\documentclass[aps,pre,reprint,groupedaddress,amssymb,showkeys]{revtex4-2}

\makeatletter
\def\set@curr@file#1{%
	\begingroup
	\escapechar\m@ne
	\xdef\@curr@file{\expandafter\string\csname #1\endcsname}%
	\endgroup
}
\def\quote@name#1{"\quote@@name#1\@gobble""}
\def\quote@@name#1"{#1\quote@@name}
\def\unquote@name#1{\quote@@name#1\@gobble"}
\makeatother
\usepackage{graphicx}
\usepackage{dcolumn}
\usepackage{bm}
\usepackage{amsmath}
\usepackage{xcolor}
\usepackage{hyperref}
\usepackage{nicefrac}
\usepackage{caption}
\usepackage{subcaption}
\usepackage{epstopdf}
\usepackage{tikz}
\usepackage{caption}
\usepackage{xcolor, soul} 
\makeatletter 
\newcount\SOUL@minus
\makeatother  
\sethlcolor{yellow}
\setstcolor{red} 

\usepackage{ulem} 

\usepackage{multirow} 

\begin{document}
	
	\title{Geometry of commutes in the universality of percolating traffic flows}
	
	\author{Sasan Ebrahimabadi$^1$}
	\author{Ali Hosseiny$^2$}\email{al_hosseiny@sbu.ac.ir} 
	\author{Jingfang Fan$^3$} 
	\author{Abbas Ali Saberi$^{4,5}$}\email{ab.saberi@ut.ac.ir \& saberi@pks.mpg.de}  
	\affiliation{$^1$School of Computer Science, Carleton University, Ottawa ON K1S-5B6, Canada}
	\affiliation{$^2$Department of Physics, Shahid Beheshti University, Evin, Tehran, 1983969411, Iran}
	\affiliation{$^3$School of Systems Science, Beijing Normal University, 100875 Beijing, China}
	\affiliation{$^4$Department of Physics, University of Tehran, P. O. Box 14395-547, Tehran, Iran}
	\affiliation{$^5$Max Planck Institute for the Physics of Complex Systems, 01187 Dresden, Germany}

	\date{\today}
	
	\begin{abstract}
		Traffic congestion is a major problem in megacities which increases vehicle emissions and degrades ambient air quality. Various models have been developed to address the universal features of traffic jams. These models range from micro car-following models to macro collective dynamic models. Here, we study the macrostructure of congested traffic influenced by the complex geometry of the commute. Our main focus is on the dynamics of traffic patterns in Paris, and Los Angeles each with distinct urban structures. We analyze the complexity of the giant traffic clusters based on a percolation framework during rush hours in the mornings, evenings, and holidays. We uncover that the universality described by several critical exponents of traffic patterns is highly correlated with the geometry of commute and the underlying urban structure. Our findings might have broad implications for developing a greener, healthier, and more sustainable future city.
	\end{abstract}
	
	\maketitle

	\textit{Introduction.}---Studying the behavior of vehicular traffic has attracted the attention of researchers for a long time. Excessive use of vehicles could bring about various problems, one of which is elevating the congestion. People are dealing with congestion almost every day, and it brings negative effects on their lives. A thorough analysis \cite{schrank20152015} of traffic situations in $471$ urban areas across the United States has revealed that travel delays due to traffic congestion pushed drivers to waste more than $3$ billion gallons of fuel and kept travelers stuck in their cars for about $7$ billion additional hours---$42$ hours per rush-hour commuter. These all translate to a total nationwide price loss of $160$ billion during traffic congestion or $960$ per commute. In addition, the ''2019 Urban Mobility Report'' remarked that traffic delay was equivalent to nearly $7$ full working days of motorists in $2017$. The negative cost of this delay could cause a loss of over $1000$ dollars \cite{schrank2019urban}.

	The other problem that emerges from congestion is the emission of pollutants into the air. Studies on the source of fine particulate matter in different areas in the United States indicated that motor vehicles are one of the primary contributing factors to air pollution and consequently to global warming \cite{rizzo2007fine,hammond2008sources}.
	The stop-and-go is a common phenomenon that occurs during congestion. Consecutive acceleration and deceleration in stop-and-go will lead to extensive burning and consumption of fuel and consequent air pollution. Releasing harmful fine particulates is highly related to mortality risk. Lung cancer and cardiovascular mortality are increased as a result of high amounts of pollutants in the air \cite{yeo2009understanding,laval2010mechanism,laden2006reduction,pope2006health,stern1977air,matsumoto2014comparative}. These effects have led scientists to find an optimal way to mitigate the congestion by exploring different aspects of traffic.

	Greenshields et al. were pioneers in measuring the speed, capacity, traffic flow, and density by taking photos \cite{turner201175,greenshields1934photographic,greenshields1935study}. Researchers started to probe into the behavior of traffic both from macro and micro points of view. In this regard, different approaches were proposed to solve the problems that traffic carries (See \cite{helbing2002micro,helbing2001traffic} and references therein).

	Totally Asymmetric Simple Exclusion Processes (TASEP), Cellular Automaton (CA), and Car Following models are classified in the microscopic approaches in which monitoring the occupancy of each cell is essential \cite{helbing2001traffic,krapivsky2010kinetic,de1999exact,gipps1981behavioural,nagel1992cellular,emmerich1997improved,maerivoet2005cellular}. The impact of roundabouts, crossroads, sudden acceleration, and deceleration, overtaking on multiple roads are the most frequent questions in this approach \cite{pronina2006asymmetric,tang2007new,foulaadvand2016phase,foulaadvand2007asymmetric,wang2002modeling,helbing2001traffic}.
	On the other hand, macroscopic models often investigate the global impact of traffic throughout the city. One of the most well-known macroscopic models was proposed by Ligthill-Whitham and Richards (LWR) \cite{helbing2001traffic,lighthill1955kinematic,richards1956shock} where the authors have utilized a first-order partial differential equation to explain the dynamic of traffic.
	
	\begin{figure*}{
			\centering
			\textbf{a} \includegraphics[width=0.3\textwidth]{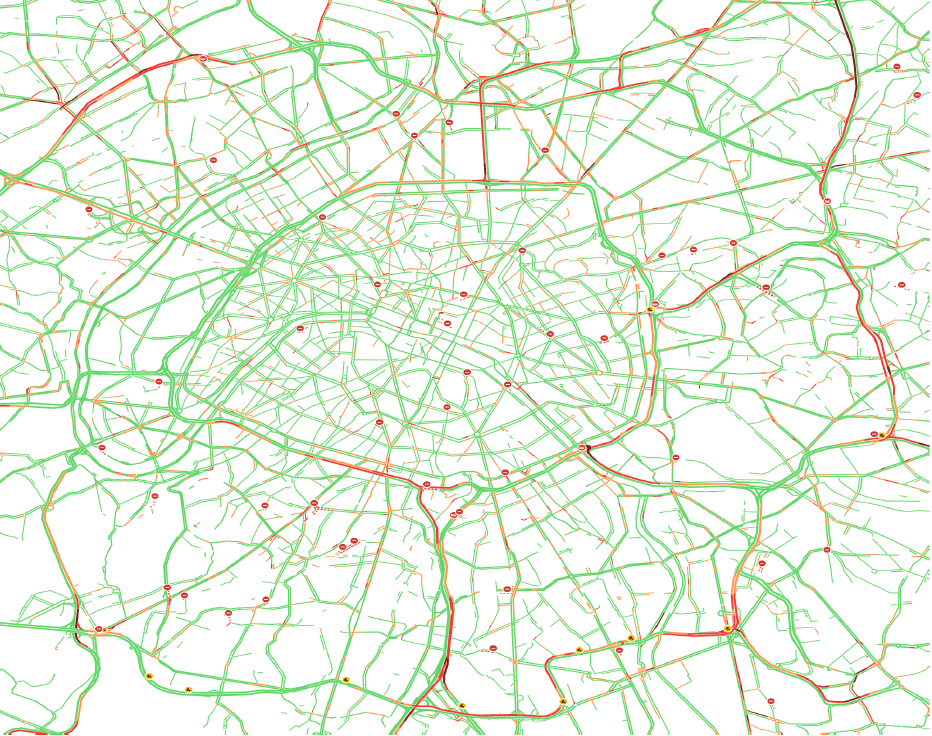}\vspace{0.25cm}
			\textbf{b} \includegraphics[width=0.3\textwidth]{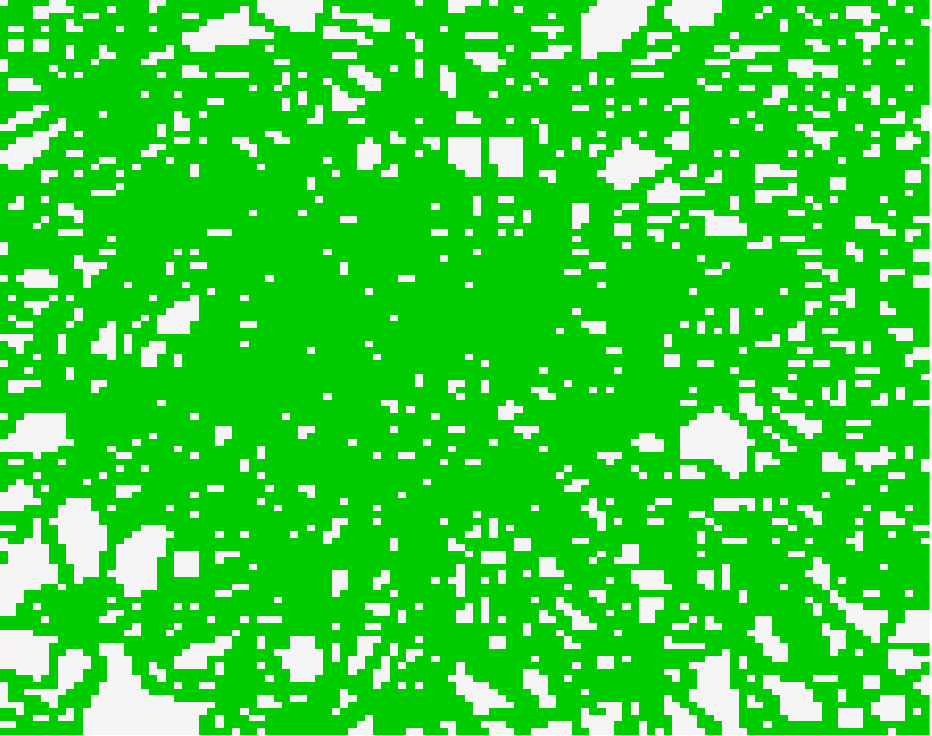}\\
			\vspace{0.25cm}
			\textbf{c} \includegraphics[width=.3\textwidth]{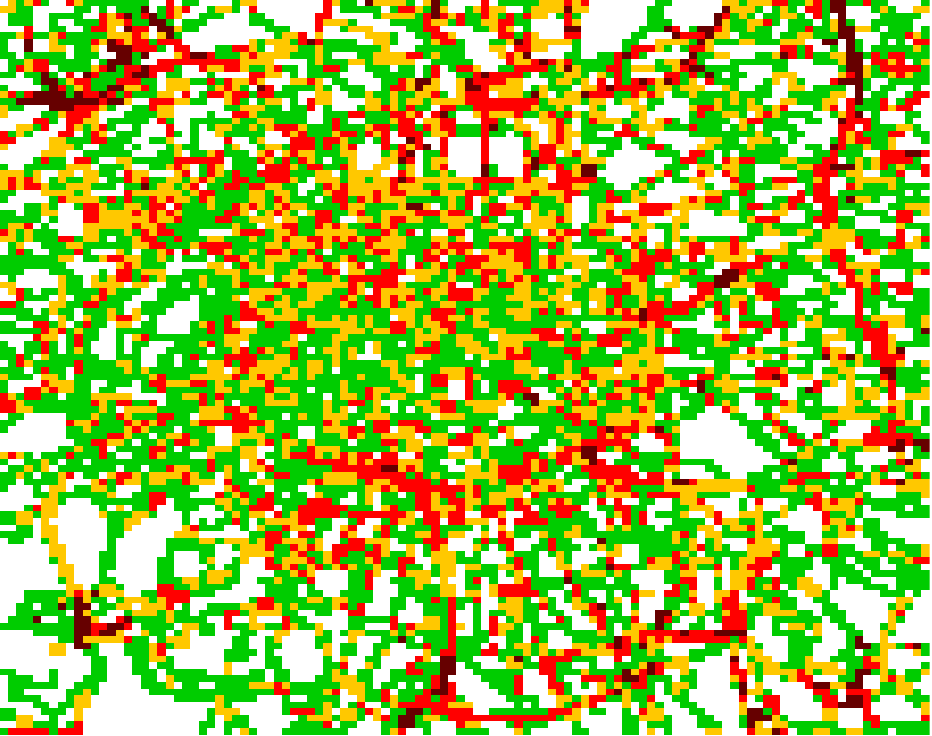} 
			\vspace{0.25cm}
			\textbf{d} \includegraphics[width=.3\textwidth]{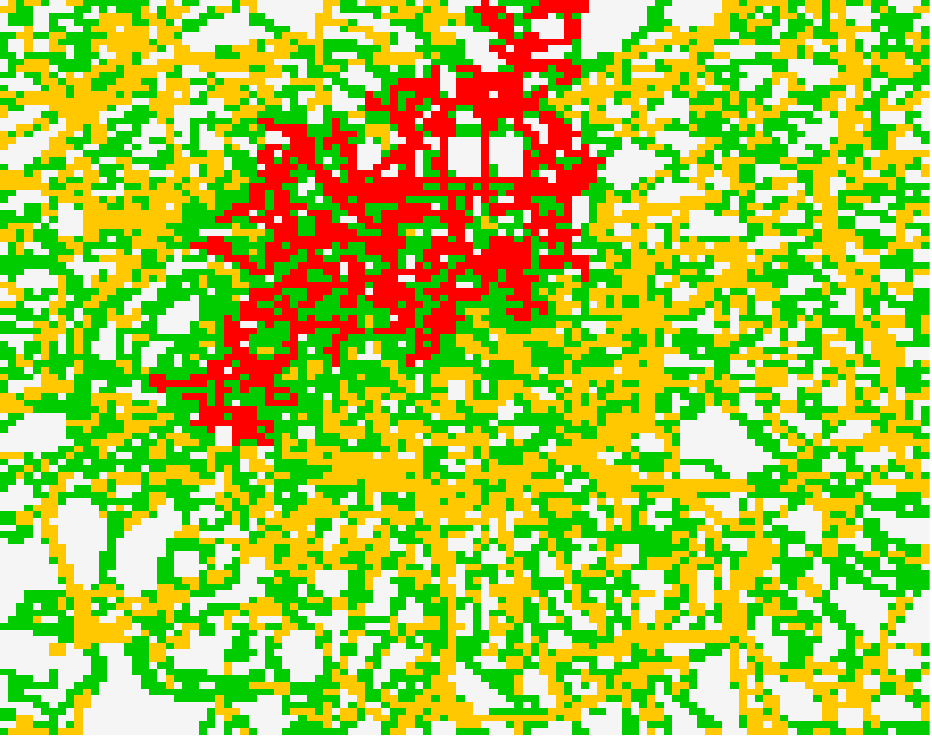}\\
			\caption{{\bf Mapping traffic pattern in Paris to percolation.} \textbf{a}. The regular Google map. Different colors indicate the intensity of traffic on roads. \textbf{b}. The map is embedded onto a $112\times112$ grid. In the first step, we indicate cells in which at least one of the pixels belongs to a road. A porous lattice appears in green cells. We will analyze the percolation of traffic on this porous lattice. \textbf{c}. Cells are colored by the traffic report of Google. If only one pixel within a cell is assigned by either orange, red, or dark red, we then consider the whole cell as a traffic cell. \textbf{d}. Percolation of traffic congestion in the porous lattice. Traffic clusters are displayed in yellow. The red cluster indicates the giant cluster.  }
			\label{figone}}
	\end{figure*}

	Propagation of congestion throughout the city and identifying the traffic zone in the city is one of the frequent questions in macroscopic models. Daganzo \cite{daganzo1993cell,daganzo1995cell} has introduced the cell transmission model which shows the evolution of traffic flow over complex networks. This model can investigate and predict the dynamics of traffic including nucleation, spanning, and the emergence of queues due to congestion.

	Percolation theory \cite{saberi2015recent} is a useful tool for studying the organization of global traffic flow on a lattice model \cite{cohen2010complex, bunde2012fractals}. Recent studies on macroscopic models have utilized percolation theory to disclose the propagation of traffic in urban areas \cite{li2015percolation,zeng2019switch}. In this geometric approach, locally congested roads will form traffic clusters that gradually grow over time and eventually merge into one giant cluster. The critical point or the percolation threshold $p_c$ coincides with the emergence of the giant cluster that is of great practical importance. The percolation problem is one of the most important universality classes in the critical phenomena characterized by a set of genuine critical exponents. Scaling relations ensure that all critical exponents can be obtained from only two independent ones e.g., the fractal dimension $D_f$ of the giant cluster and the Fisher exponent $\tau$ governing the cluster size distribution at criticality.
	Determination of the universality class of traffic clusters can indeed provide insights into the underlying mechanism and influential parameters that shape the propagation of traffic jams over time \cite{dorogovtsev2008critical,cohen2010complex,barzel2013universality}.
	
	\begin{figure}[]
		{   \centering
			\includegraphics[width=0.8\columnwidth]{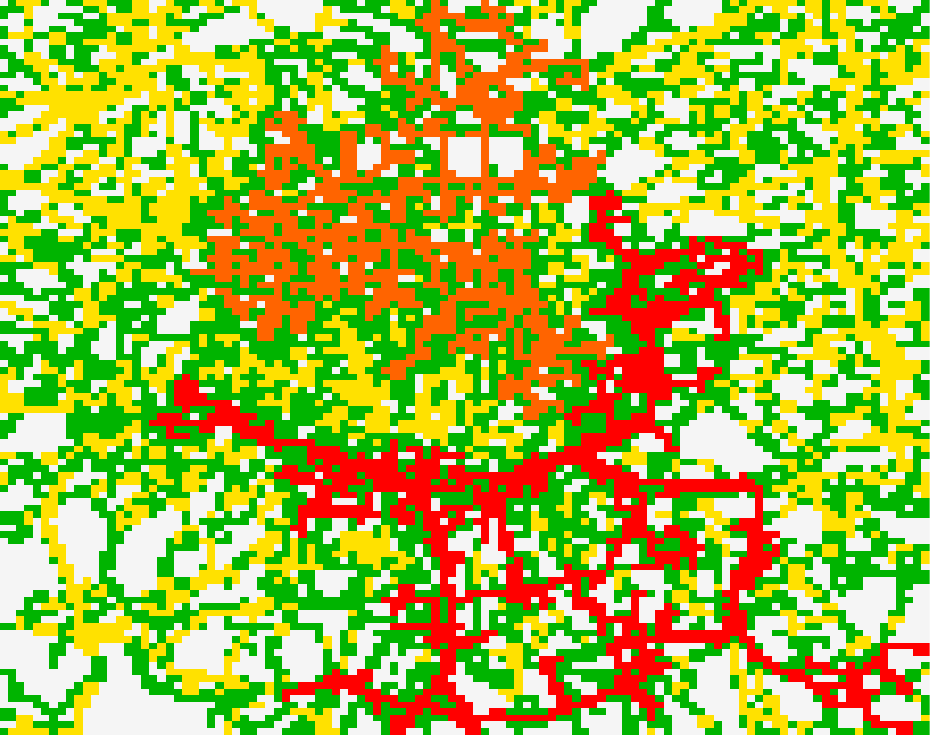}\\
			\caption{{\bf Giant cluster of traffic jam in Paris.} The red cluster represents the giant cluster at the percolation critical point in Paris in the morning of a working day. The orange cluster represents the second largest cluster. }
			\label{morningcluster}}
	\end{figure}
	
	\begin{figure*}[]{
			\centering
			\includegraphics[width=0.8\textwidth]{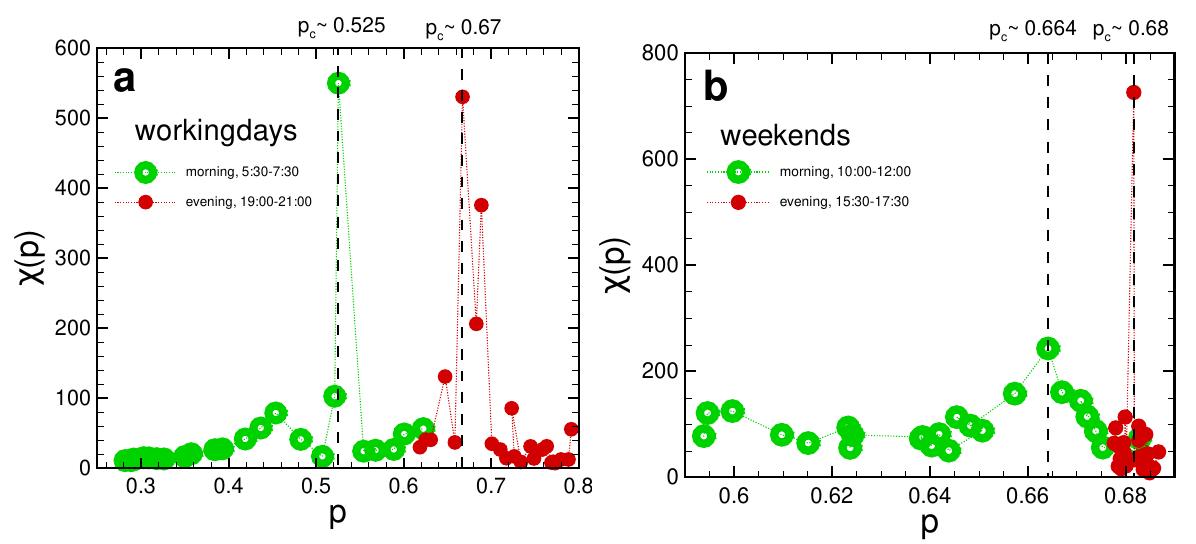}
			\caption{{\bf Mean cluster size in Paris.} The mean cluster size $\chi(p)$ as a function of the traffic rate $p$ in the working days (\textbf{a}) and weekends (\textbf{b}) in Paris. The vertical dashed lines show the location of maximums at the critical traffic rate $p_c$. }
			\label{figmeanclustersfridaymorning}}
	\end{figure*}

	Here, we study the structure of traffic clusters in Los Angeles and Paris by using the concepts of percolation theory. We compare their properties in the morning and the afternoon/evening rush hours of weekdays and weekends. The critical question is whether the percolation properties of traffic clusters in these cities with totally different road network structures are influenced by the geometry of the commutes. 

	\textit{Results.}---We first embed the Paris map into a $112\times112$ grid in our analysis. Snapshots are taken from Google Maps for live traffic patterns every $5$ minute in $14$ days during June and November $2018$. Roads in Google Maps carry four possible colors: green, orange, red, and dark red. Colorless regions in the map are related to places other than roads, which we will not consider in this study (Fig.  \ref{figone}). The green stands for a road in the traffic-free mode. In other cases, however, traffic is involved with the increasing intensity from orange to dark red.  
	This setting provides a network of roads of Paris and Los Angeles embedded in a discretized two-dimensional space (Fig.  \ref{figone}). 
	\begin{figure}{
			\centering
			\includegraphics[width=.8\columnwidth]{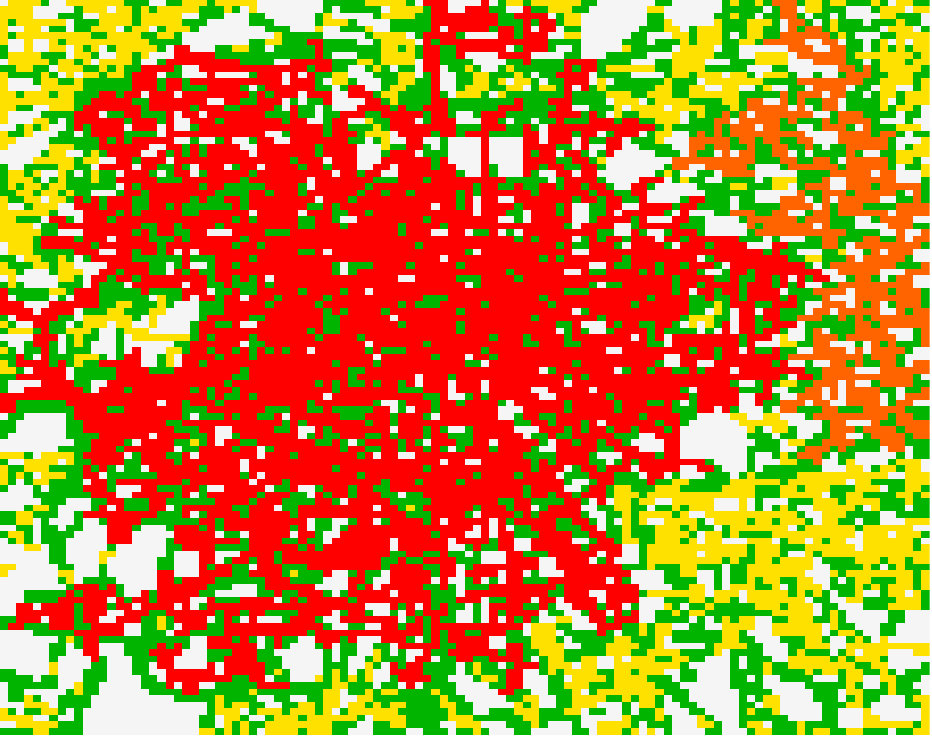}\\
			\caption{{\bf Giant cluster of traffic jam in weekends.} The red cluster represents the giant cluster at the percolation critical point in Paris on a weekend. The orange cluster represents the second largest cluster. 	}
			\label{weekendcluster}}
	\end{figure}

	\begin{figure*}{
			\centering
			\includegraphics[width=0.6\textwidth]{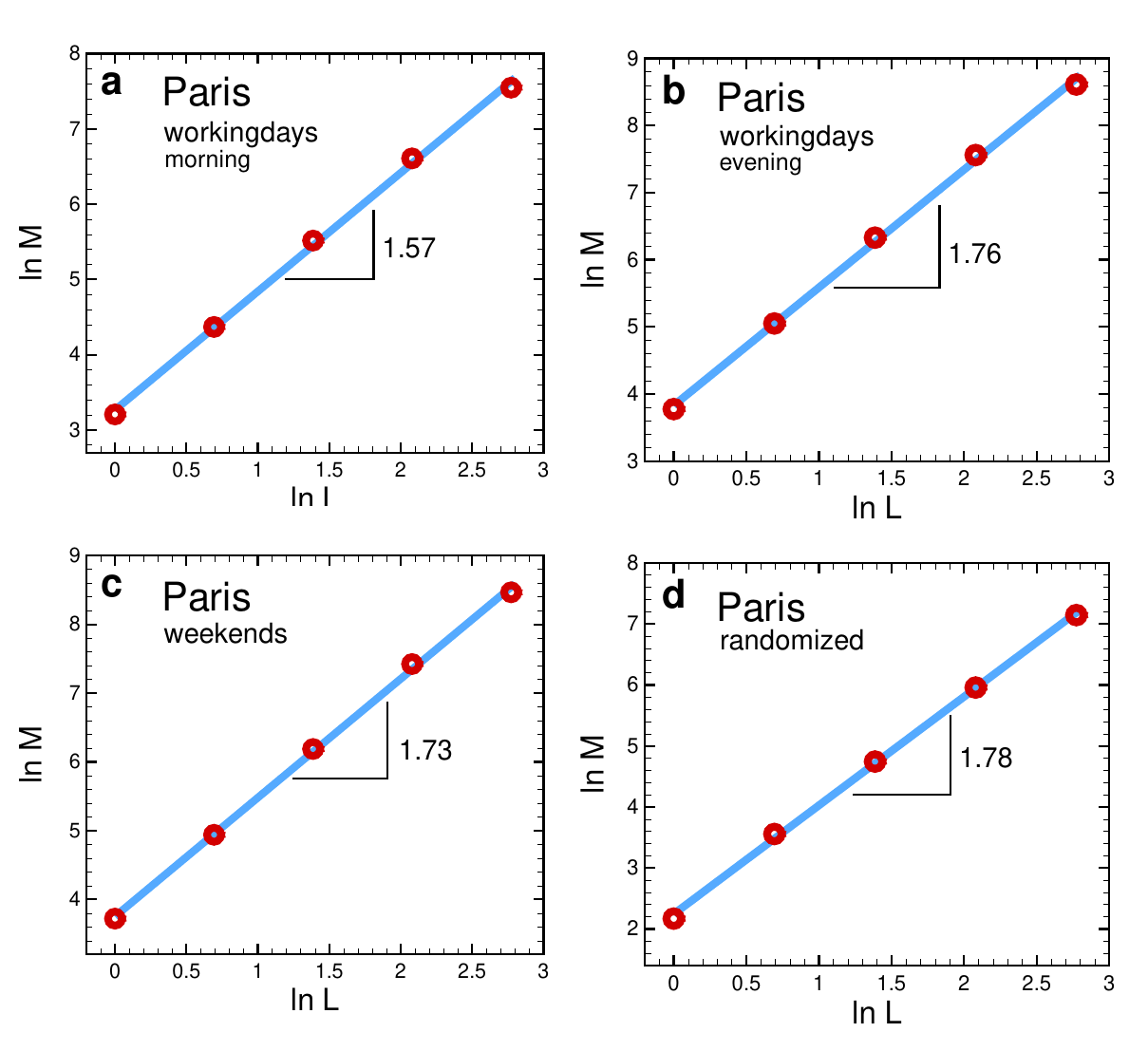}
			\caption{{\bf Fractal property of the giant cluster:} $\log$-$\log$ plots of the average mass $M$ of the giant cluster of a traffic pattern in a window of size $L$ in the mornings of the working days (\textbf{a}), evenings of the working days (\textbf{b}), weekends (\textbf{c}) in Paris. (\textbf{d}) Implementation of random percolation model on the city road geometry. The slopes give the fractal dimension of the giant cluster at the critical thresholds. The solid lines are the best fits to our data with $R^2>0.99$. The error bars (based on standard error) are the same size as the symbols. }
			\label{boxcounting}}
	\end{figure*}

	\textit{Materials and Methods.}---Google Maps has a feature called “Google Traffic” that provides accurate real-time traffic information to online users. The information is graphically coded in four different colors: green for the traffic-free mode, orange for moderate traffic, red for high traffic, and dark red for traffic congestion (Fig. \ref{figone}). 
	We captured the live traffic patterns of Paris and Los Angeles in $5$ minute intervals for two weeks. There are different features in the snapshots that we captured (e.g. names of roads, rivers, Parks, etc.)  but we only used the graphical information for the traffic report.  
	
	In the first step, we use image processing to convert the RGB matrix (which is a 3D matrix ) into a 2D matrix with values ranging from 1 to 4 each for colors from green to dark red, respectively (Green=1, orange=2, red=3, and dark red=4). We have then embedded the matrix information onto a square lattice of size $1792\times1792$. Since the cell size on our original lattice was less than a typical road width, we have used a coarse-grained lattice by merging each $16\times16$ cell into one larger cell. Thus our initial lattice of size $1792\times1792$ shrank into a lattice of size $112\times112$. 
	
	\textit{Models}. we used the site-percolation model to investigate traffic propagation and dynamics throughout Paris and Los Angeles. We consider a site (or cell) to be congested if its attributed value is higher than $1$. This means that all sites with colors in orange, red, and dark red are congested. We define $q = 2$ as a threshold for each cell in the matrix $A_{ij}$ i.e.,
	\begin{equation}
		A_{ij}=\begin{cases} 1 \qquad \text{if} \qquad \qquad & A_{ij} \geq q \\
			0 \qquad \text{if} \qquad \qquad & A_{ij} < q \end{cases}
	\end{equation}

	\textit{Map onto a percolation problem.} 
	Fig. \ref{figone}\textbf{a} represents a typical snapshot of the traffic pattern in Paris. Fig. \ref{figone}\textbf{b} shows the porous lattice of the road network on which our percolation analysis is carried out. Fig \ref{figone}\textbf{c} is an example of the transformation of the original 3D RGB Google Maps traffic pattern into the two-dimensional $112\times112$ lattice. Each lattice cell may contain several colored pixels in the original map. We use the max-pooling method which assigns the darkest color of pixels to the cell. We then consider all cells with different colors as occupied cells in percolation and represent them with orange. The nearest neighbors of occupied sites on the lattice form a connected cluster, and the number of occupied sites in this cluster defines its size $s$. The giant cluster with the largest size at the given traffic rate is shown in red in Fig \ref{figone}\textbf{d}. The traffic rate is related to the occupation probability $p$ that measures the ratio of the total number of occupied sites to the total number of green sites i.e., all sites that belong to the road network. Once $p$ reaches a critical threshold $p_c$ i.e., the large-scale traffic is jammed and the giant cluster spans across the lattice. 
	
	To capture these critical points, we first measure the mean cluster size (analogous
	to the susceptibility of the system),  $\chi(p)$ defined by
	\begin{equation}
		\chi(p)=\Sigma'_{s}s^2n_s(p)/\Sigma'_{s}sn_s(p), 
		\label{eqaverage}
	\end{equation}
	where $n_s(p)$ denotes the average number of clusters of size $s$ at each traffic rate $p$, and the prime on the sums indicates the exclusion of the largest cluster in each measurement.

	Fig. \ref{morningcluster} illustrates an example of the traffic pattern at the critical rate in the morning of a working day in Paris. The red cluster is the largest cluster and the orange one is the second largest cluster.

	Fig. \ref{figmeanclustersfridaymorning}\textbf{a} represents the mean cluster size of traffic pattern in both morning (green open circles) and evening (red solid circles) on a working day in Paris. The divergence in $\chi(p)$ signals the critical traffic rate at the onset of large-scale traffic congestion. Remarkably, the critical traffic rate in the morning is smaller than the critical rate in the evening. We find that the average traffic rate in the morning for the observed period is $0.525\pm0.05$ and for the evening is $0.67\pm0.02$. This means that the morning of working days in Paris carries a lower capacity of traffic flow which can be caused by the common and localized destinations of vehicles in the morning and thus common roads. Fig. \ref{figmeanclustersfridaymorning}\textbf{b} shows the percolation of traffic flow during a weekend in Paris. As it can be seen, the critical traffic rate $p_c$ in the morning ($\sim 0.664\pm 0.02$) and evening ($\sim 0.68\pm 0.02$) is very close to that of the evening of the working days ($\sim0.67\pm0.02$). In the evenings and weekends, the destinations are chosen by the people which are well distributed all over the city which provides more variety of routes for vehicles and as a result the traffic load of the city increases in this period of time. This difference is also well represented schematically in the giant cluster of the traffic pattern for weekday morning in Fig. \ref{morningcluster} which is a highly porous and low-density fractal structure versus a more dense structure of the giant cluster observed in the weekend shown in Fig. \ref{weekendcluster}.

	Such observations prompt us to investigate whether the difference between weekend and weekday evening critical thresholds versus weekday morning thresholds is indicative of fundamental differences in traffic patterns from the perspective of critical complex systems. The universality allows to grouping of microscopically quite different physical models and phenomena with the same behavior near criticality into universality classes characterized by a set of critical exponents related to the broad symmetry groups. In
	percolation theory, the critical exponents, unlike the critical threshold, do not depend on the microscopic details of the underlying lattice but only on the Euclidean dimension $d$ and the dimensionality of the order parameter.
	
	\begin{figure}{
			\centering
			\includegraphics[width=0.4\textwidth]{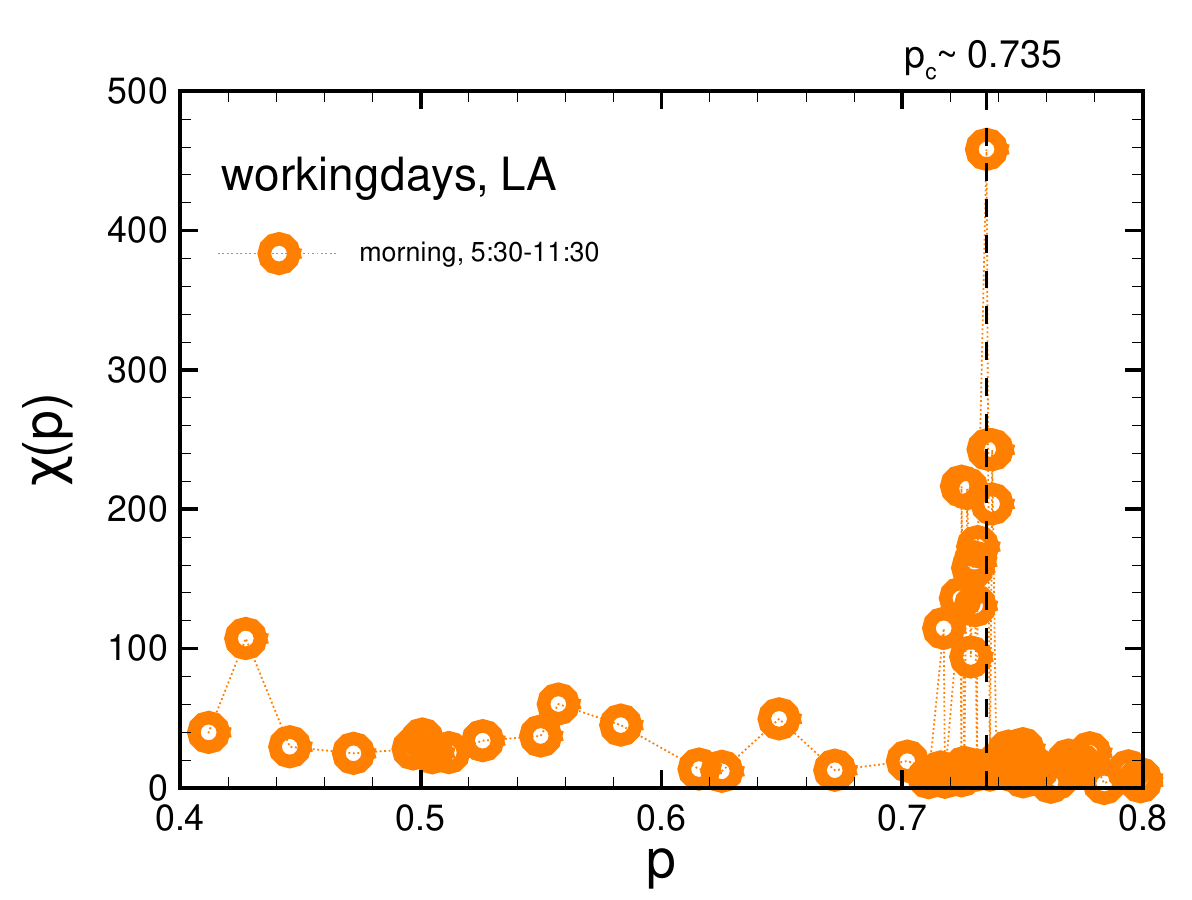} 
			\caption{{\bf Mean cluster size in LA.} The mean cluster size $\chi(p)$ as a function of the traffic rate $p$ in the working days in LA. The vertical dashed line shows the location of maximums at the critical traffic rate $p_c$. }
			\label{lapercolation}}
	\end{figure}

	To evaluate the universality class of the traffic patterns at the critical rates, we first measure the fractal dimension $D_f$ of the giant clusters which also implies
	the emergence of self-similarity in the geometric feature of the percolation clusters. If the giant cluster is scale invariant then it requires that the mean mass
	$M$ of the cluster within the window of length $L$ would increase as a power-law with size, i.e.
	\begin{equation}
		M(L)\propto L^{D_f}.
		\label{eq2}
	\end{equation}
	The results of our fractal analysis based on examination of the scaling relation Eq. \ref{eq2} are illustrated in Fig. \ref{boxcounting} where we have shown $\log$-$\log$ plots of the average mass of the giant cluster within a window of linear size $L$. The slope of the linear fit to our data gives the best estimation for the fractal dimension. We find $D_f=1.57\pm0.05$ for the giant cluster of traffic jam in the morning of the working days (Fig. \ref{boxcounting}\textbf{a}), while we find higher values $1.76\pm0.05$ (Fig. \ref{boxcounting}\textbf{b}) and $1.73\pm0.05$ (Fig. \ref{boxcounting}\textbf{c}) for the evening of the working days and the weekends, respectively, in agreement with our earlier graphical observations from Figs. \ref{morningcluster} and \ref{weekendcluster}. 
	
	In order to further elucidate the nature of the universality classes observed at different times in Paris, we examine the implementation of the random percolation model on the geometry of the road network in Paris. To this aim, we randomly choose a site on the roads and occupy it with a vehicle. We find that the model shows a critical behavior at a critical rate $p_c=0.69\pm 0.04$ and the fractal dimension of the giant cluster is estimated to be $D_f=1.78\pm 0.03$ (Fig. \ref{boxcounting}\textbf{d}). Remarkably, this random percolation model agrees well with our observations of the real traffic situations in Paris during the evenings of the working days and the weekends. This finding is in accord with our previous justification of the random distribution of vehicle destinations during the evening of working days and weekends.

	\begin{table*}
		\centering
		\caption{Percolation threshold $p_c$, fractal dimension $D_f$, and the Fisher exponent $\tau$ for the considered cities and models. \\}
		\label{table1}
		\begin{tabular}{|c|c|c|c|c|}
			\hline
			\multirow{2}{*}{City/Model} &\multirow{2}{*}{$p_c$} &\multirow{2}{*}{$D_f$}& \multicolumn{2}{c|}{$\tau $} \\
			
			\cline{4-5}
			& & & exponent & p-value \\
			\hline
			working day mornings in Paris & $0.525\pm 0.05$ & $1.57\pm 0.05$ &  $2.33\pm0.04$& $0.25$\\
			\hline
			working day evenings in Paris & $0.67\pm0.02$ & $1.76\pm 0.05$ &  $2.17\pm0.04$ & $0.70$\\
			\hline
			weekends in Paris & $0.68\pm0.02$ & $1.73\pm 0.05$ & $2.16\pm0.05$ & $0.21$ \\
			\hline
			random percolation on Paris& $0.69 \pm 0.04$ & $1.78\pm 0.03$ & $-$ & $-$\\
			\hline
			working day mornings in LA& $0.735\pm0.05$ & $1.77\pm0.05 $ & $2.16\pm0.04$ & $0.63$ \\
			\hline
			random percolation on LA & $0.70 \pm 0.04$ & $1.70\pm 0.04 $ & $-$ & $-$ \\
			\hline
			2D site percolation on square lattice & $0.592 \pm 0.003$ & $91/48 $ & $187/91$ & $-$\\
			\hline
		\end{tabular}\label{tableparis}
	\end{table*}
	
\begin{figure}{
			\centering
			\includegraphics[width=0.45\textwidth]{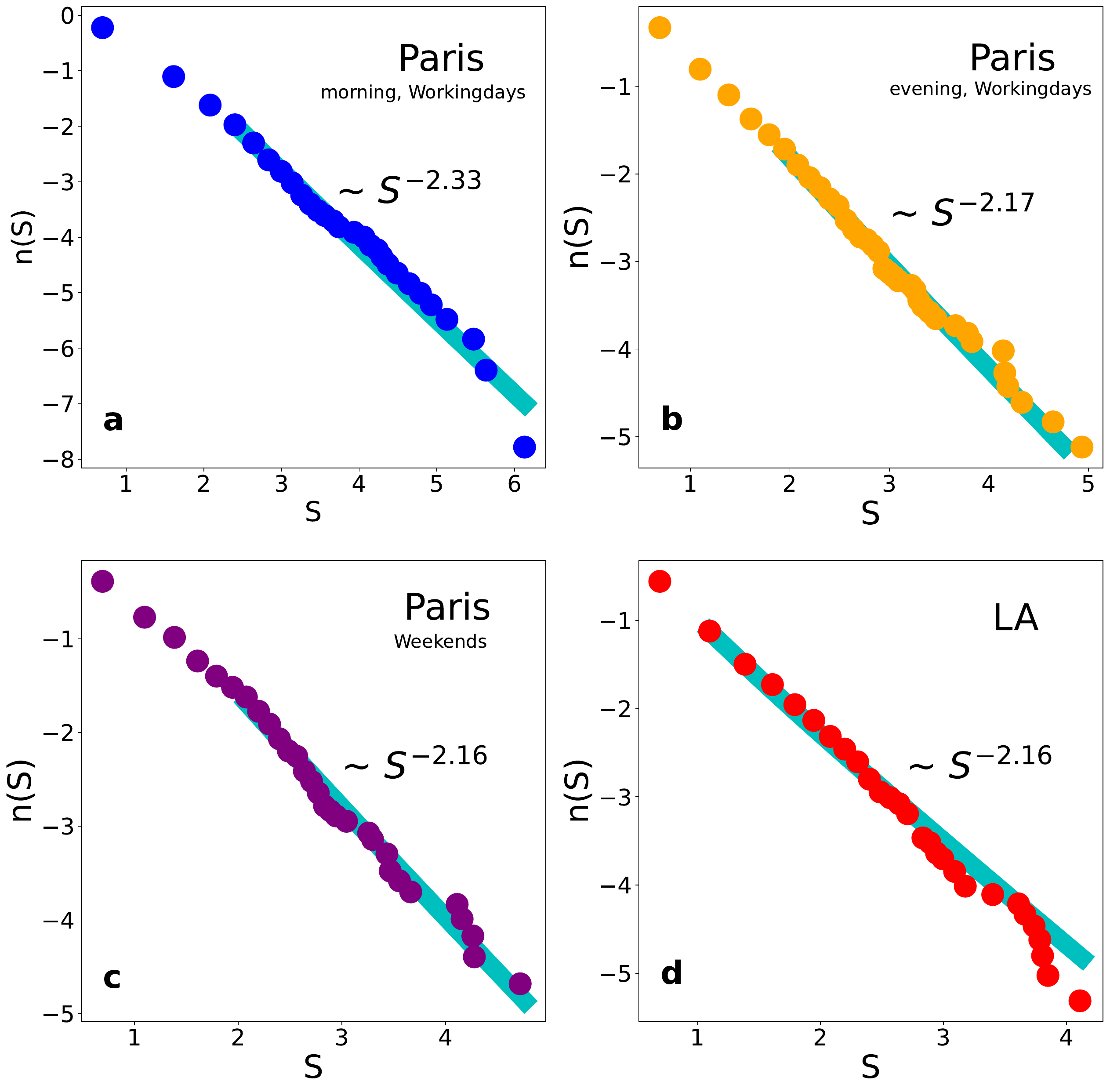} 
			\caption{{\bf  The Fisher critical exponent $\tau$ at the percolation threshold.} (a) Working days in the morning in Paris, (b) Working days in the evening in Paris, (c) Weekends in Paris, and (d) LA.}}
		\label{Figure-tau-exponent}
	\end{figure}

	We have also performed a similar analysis on the statistical behavior of the traffic pattern in the city of Los Angeles (LA). The urban structure of LA and the road network are completely different from the city of Paris, and this helps to better understand the effect of commute geometry on the universality and the traffic threshold. Figure \ref{lapercolation} shows the mean cluster size as a function of traffic rate in the morning of LA. The global maximum in $\chi$ happens at $p_c=0.735\pm0.05$ which is close to that of the random percolation model on Paris as well as our observations on the evenings and the weekends in Paris. To examine its universal properties at the criticality, we have measured an independent critical exponent i.e., the Fisher exponent $\tau$ shown in Fig. \ref{Figure-tau-exponent} using the following scaling relation at the critical threshold
	\begin{equation}  
		n(s)\propto s^{-\tau}.
		\label{eq1}
	\end{equation}
	
	We use the algorithm described in \cite{clauset2009power} to evaluate the Fisher exponents. To this aim, we first estimate the lower bound of the traffic cluster size distribution where it begins following a power-law behavior. Thereafter, we used maximum likelihood estimation (MLE) to find the best scaling parameter that fits our observed data. To check the validity of our measured scaling parameter, we used the goodness of fit test which would generate a p-value. We measure the p-value by using the Kolmogorov–Smirnov (KS) test to calculate the distance between the empirical data and the hypothesized power-law model. Afterward, we create a large number of synthetic data with the given scaling parameter and lower bound and then calculate the distance between the model and each synthetic data set by using KS statistics. We can define the p-value once we calculate the fraction of times that the KS result is larger than that of real data. It is reasonable to say that a distribution follows a power-law behavior if the p-value is greater than $\sim0.1$ (for a detailed explanation of the method see \cite{clauset2009power}).

	By employing the aforementioned method, we find the Fisher exponent $\tau\sim2.33$ for the mornings of the working days in Paris and almost the same exponents around $\tau\sim2.16$ for the evenings and the weekends of Paris and LA (Table \ref{tableparis}). This shows the crucial role that is played by the geometry of commutes in shaping the universal and characteristic properties of traffic patterns in megacities. We have also carried out random percolation analysis on the LA road network and found results close to those obtained in Paris. All measured thresholds and exponents are summarized in Table \ref{table1}. These all support our conclusion that the commute geometry leaves its footprints on the traffic patterns. 
	

	\textit{Conclusion.}---In classical critical phenomena, the universal features are independent of microscopic details and only dependent on the dimensionality and the underlying symmetries. In a given support dimension, changing the universality class requires manipulating interactions in a relevant manner. In the percolation model that we used in the analysis of the traffic patterns of Paris, the universality class in the mornings of working days is distinctly different from the weekends and evenings of working days. The only important difference in these two situations is the change in the distribution of supply and demand at the city level, which seems to have a much more significant effect than what was thought. Because even when we look at the dynamics of the traffic pattern in the city of Los Angeles, with a different city structure and geometry from Paris, the universality class is like the weekends in Paris, which is also indistinguishable from a case where vehicles are randomly distributed in the city. These observations are very promising because our results suggest that without manipulating the road network and urban structure (urban geometry) which is very costly if not impossible, the critical traffic rate of the city can be significantly increased by changing the distribution of supply and demand sources in the city. This means that the universality class of the traffic model is manageable based on the random percolation model, which seems to provide an optimal condition for city traffic.
	
	\textit{Acknowledgement}
	The authors thank Aaron Clauset for providing and openly sharing the codes for calculating the exponent of power law distributions.

\end{document}